\pdfoutput=1

\documentclass[11pt]{article}

\usepackage[]{ACL2023}

\usepackage{times}
\usepackage{latexsym}

\usepackage[T1]{fontenc}

\usepackage[utf8]{inputenc}

\usepackage{microtype}

\usepackage{inconsolata}

\usepackage{multirow}
\usepackage{graphicx}
\usepackage{bm}
\usepackage{times}
\usepackage{latexsym}
\usepackage{amsfonts}
\usepackage{bm}
\usepackage{placeins}
\usepackage[capitalise,noabbrev]{cleveref}

\title{What Food Do We Tweet about on a Rainy Day?}

\author{Maija Kāle \\
  Faculty of Computing \\
  University of Latvia \\
  \texttt{\{name.surname\}@lu.lv} \\\And
  Matīss Rikters \\
  National Institute of Advanced \\
  Industrial Science and Technology \\
  \texttt{\{name.surname\}@aist.go.jp} \\}

\begin{document}
\maketitle
\begin{abstract}
Food choice is a complex phenomenon shaped by factors such as taste, ambience, culture or weather. 
In this paper, we explore food-related tweeting in different weather conditions. We inspect a Latvian food tweet dataset spanning the past decade in conjunction with a weather observation dataset consisting of average temperature, precipitation, and other phenomena. We find which weather conditions lead to specific food information sharing; automatically classify tweet sentiment and discuss how it changes depending on the weather. This research contributes to the growing area of large-scale social network data understanding of food consumers' choices and perceptions.
\end{abstract}

\section{Introduction}
\label{intro}

This paper focuses on the relationship between food sentiment and weather using the previously collected Latvian Twitter Eater Corpus (LTEC \cite{SprogisRikters2020BalticHLT}). We seek to answer (1) is there a correlation between food sentiment and weather experienced at the time of tweeting and (2) what are the differences in the term frequencies of food mentioned depending on the weather. The rationale for this paper is to contribute to deeper understanding of human-food relationship, in particular in relation to weather data. We believe that with more nuanced knowledge of human-food relationships and factors influencing them, we can provide valuable inputs for public health policy makers when they develop their strategies and nudge consumers to choose more healthy options of food.
Weather people - this is a term that Bakhshi \cite{WeaPpl} used to explain our dependence on the weather regarding food choices and satisfaction with food. While the weather is known to alter consumers' mood significantly and consequently their behaviour \cite{Bujisic2019}, there have been surprisingly few studies that illustrate weather's impact on food perception and food choices, except some that have used online and offline restaurant reviews as a proxy of measuring it \cite{Bakhshi2014,Bujisic2019}. They find that weather impacts both the frequency of the feedback that food consumers provide, as well as its content. Typically, sunny and pleasant weather leads to more frequent and more positive feedback, since low levels of humidity and high levels of sunlight are associated with high mood. At the same time, reviews written on rainy or snowy days, namely days with precipitation, tend to have lower ratings.
Instead of analysing restaurant reviews, we focus on Twitter, where food represents one of the key themes discussed, providing us with spontaneous reactions, which is a unique feature when compared to other data collection methods like reviews or food diaries \cite{PUERTA2020103993}. Our analysis of the LTEC provides a food-related set of discussions that we can correlate with weather data, leading to the following research inquiries: 1) is there a correlation between food tweet sentiment and the weather that the tweet authors are experiencing at the time of tweeting? 2) what are the differences in terms of frequencies of what food is mentioned in tweets depending on weather? 
One of the reasons, why there are few weather-food choice related studies, is the lack of data - we do not have access to retailers' food sales data that could be correlated with the weather data. Instead, we are focusing how food is represented in social media - in particular Twitter, assuming that tweet is an appropriate proxy to measure sentiment related to food consumption. By analysing weather-related dynamics in LTEC, we contribute to the research field that links food and mood, adding weather impact on the mood.

\section{Related Work}
\label{sec:related}

Food consumption is a complex process that is impacted interchangeably by various endogenous factors, such as taste, quality, texture, colour and others, as well as exogenous or external factors ranging from demography, educational level, time of the day, weather, the ambience where it is consumed and others \cite{velasco2021gastrophysics,Bujisic2019}.
Mood is the determining factor in food choice, where good mood is associated with healthier food choices and bad mood with less healthy food choices \cite{Spence2021}. Food choice is also seasonally patterned in particular in areas with more seasonal climate in terms of temperature. Even though most of our modern lives are spent indoors, weather and climate conditions still impact our food preferences and consumption \cite{SpenceSeason}. While seasonal food consumption patterns are culture-based and differ in various geographical regions, weather-related preferences seem universal. Sunny and moderate temperature-wise weather leads to better mood, while more extreme weather (hot, cold, precipitation) is less pleasant and impacts mood, food consumption experiences.

A large-scale study on demographics, weather, and restaurant reviews reveals that pleasant weather impacts not only the content but also the frequency that is higher than during non-pleasant weather conditions \cite{Bakhshi2014}. This is an important indicator that a review can serve as a proxy for measuring the weather's impact on mood and, thus, the food consumption experience. Consumer comments and word-of-mouth have also been studied in relation to weather, implying that consumers' pre-consumption mood directly influences post-consumption mood, and consumers' satisfaction with the service accordingly. Pre-consumption mood, is viewed via weather conditions, where eight weather-related variables have been considered, including  visibility, rain, storm, humidity, wind speed, pressure. By including temperature, barometric pressure, and rain as variables reduces unexplained variance and improves results of the experiment. This study successfully links weather to mood and its transfer to affective experience and consumer behaviour \cite{Bujisic2019}.

Considering previous studies that prove the link of weather to mood and food perception accordingly, with our work, we aim to illustrate this link via tweet sentiment evaluation. We refine our study by looking at frequencies - what foods authors tweet more in pleasant weather and unpleasant weather conditions, mapping the weather-related food scene in Latvian language Twitter. 

\section{Case Study of Latvia}
\label{sec:about-latvia}
Latvia has four distinct seasons: winter is December to February, spring - March to May, summer - June to August, autumn - September to November. The average annual air temperature in Latvia is only +5.9°C. The warmest month is July, and the coldest months are January, February (also the snowiest). Months with the most precipitation are July and August, while the least is in February and March. The highest wind speeds are in November, December and January, and the lowest are in July and August \cite{WeaData}. Latvia provides an example of a country in the Northern hemisphere with various weather conditions to analyse from the perspective of tweeting about food.

Besides recognising weather data, Latvian national cuisine seasonality aspects should be considered. Specific foods are consumed in certain seasons in Latvia - cold soup in summer, grey peas, tangerines and gingerbread for the Christmas season \cite{Kale-et-al-2021}. This cultural context is important for understanding weather-related impact on food tweet understanding. 

Other cyclical events that are present in any modern society should also be considered. Not just weather and seasonal celebrations are cyclical in nature and correlate with the time of the year. There are other variables that correspond to the time of year that could be possible confounds, for example, school schedules, holiday seasons, election events, sport events, etc. While aware of such cyclical events, we do not highlight them here due to lack of previous research to provide us with reference data. The only study about the timeline of food related tweets in Latvia reveals that a slight decrease of food tweeting was observed on weekend evenings, and a significant one – on weekend mornings \cite{Kale-et-al-2021}. These results imply the overall differences in mood and behaviour at various times of the day/meals: people tend to be more ‘virtuous’ in mornings by choosing healthy and nutritious food, while snacking during afternoons \cite{SPENCE2021104198}. 

The nuances to consider can be categorised in individual circadian rhythms, culture/climate bound seasonality cycles, celebrations, and cyclical events. While being aware of those multiple factors, in this work we focus on weather data primarily, linking them with tweet sentiment without additional references to cyclical nature of human life.

\section{Data Collection and Processing}
\label{sec:data}

We used a combination of the LTEC for tweets and weather data exported from Meteostat\footnote{https://meteostat.net/en/place/lv/riga}. We mainly focused on tweets and weather relating to Riga, the capital of Latvia, since most tweets with location data originated there, and it was difficult to obtain detailed historical weather data for the smaller regions.

The LTEC has a total of 2.4M tweets generated by 169k users.
It has been collected over ten years following 363 eating-related keywords in Latvian. Among the tweets, 167k have location metadata specified, of which 68k were from Riga and 9k more from areas around Riga.
To further increase the number of location-related tweets, we selected all remaining tweets which mention Riga or any of its surrounding areas (Marupe, Kekava, etc.) in any valid inflected form. This added 54k tweets, totalling to 131,595.


In addition to location metadata, the LTEC provides all food items mentioned in the text and a separate subset of sentiment-annotated tweets for training sentiment analysis models. We use the 5420 annotated tweets to fine-tune a multilingual BERT \cite{devlin-etal-2019-bert} model for this task along with $\sim$20,000 sentiment-annotated Latvian tweets from other sources\footnote{https://github.com/Usprogis/Latvian-Twitter-Eater-Corpus/tree/master/sub-corpora/sentiment-analysis}. Evaluation was performed on the 743 tweet test set from LTEC and reached an accuracy of 74.06\%. We then use the model to automatically classify the location-specific tweets as positive, neutral or negative.

We could reliably obtain only data for temperature and precipitation from Meteostat, while data for snowfall was only available up to 2017, and data for wind speed and air pressure was only available from 2018 onward. 
There was no available data to trace daily sunshine, but it can be inferred from looking at precipitation, snowfall and air pressure. 

\subsection{Limitations and Assumptions}
\label{sec:limitations}

Our work has several important limitations that can be grouped into categories of 1) data availability, 2) tweet author's demographic profile, and 3) generalisation of the results. First, we could only obtain fairly superficial weather data while weather change during the same day was not considered due to lack of detail. Second, we cannot provide a demographic outlook of the usual tweet author in LTEC, and our analysis includes tweets by general digitally literate people active on Twitter. Third, considering the limitations discussed, our results are not an exact extrapolation of weather-related food perception in Latvian society. Nevertheless, our approach adds to the understanding of weather's impact on the part of the Latvian society which tweets about food.

\section{Analysis and Results}
\label{sec:results}

While the results of tweet sentiment in terms of the percentage of negative, neutral and positive tweets are largely the same for all weather conditions, we can observe considerably fewer positive tweets during windy weather and high-pressure, as shown in Table \ref{tab:emo-weather}. Surprisingly, even during low-pressure weather conditions, tweets are not necessarily dominated by negative sentiment - quite the opposite - food tweets have been related to mostly positive sentiment. It could be explained by the fact that people are tweeting about comfort food (e.g. coffee, chocolate, other) or that any food could be comforting during days of low-pressure weather conditions. This remains to be answered in a more fine-grained manual analysis.

The right part of Table \ref{tab:windy-rainy-cold-warm} shows that tea exceeds coffee during cold weather, and there is also a slight increase in tweets about chocolate in cold weather, while the frequency of ice-cream tweets doubles in warm weather. Interestingly, in hot or cold weather tweet amount about meat, cake or soup remains largely similar. While warm weather tweets include strawberries, cold weather tweets include gingerbread, which coincides with seasonal Christmas food. There are no other notable differences between warm and cold weather tweets, which leads to a conclusion that spending so much time indoors has harmonised foods tweeted about in different seasons and conditions.

A slightly different result is revealed in the left part of Table \ref{tab:windy-rainy-cold-warm}, which indicates that during windy weather, meat becomes the most popular food item, while in rainy weather, the results are similar to cold weather where tea dominates. While it is difficult to explain this, a speculation could be that wind is less visible than temperature that is frequently reported in media or precipitation that is visually noticeable before leaving the home, and, thus, without proper clothing during windy weather one might become uncomfortably cold, which in turn could lead to higher willingness to consume meat. Chocolate is twice as popular during rainy weather than during windy weather, and it could be related to a lack of sunshine during rainy weather that needs to be compensated with chocolate, while a windy day can still be sunny.

Only potatoes remain stable in terms of tweeting frequencies in any weather - warm, cold, windy or rainy. This can be explained by the fact that potatoes are part of a daily diet in Latvia and constitute the basis for energy intake.

\begin{table}[t]
    \centering
    \small
    \begin{tabular}{|l|c|c|c|c|}
    \hline
    \multicolumn{1}{|c|}{\textbf{Product}} & \textbf{Rainy} & \textbf{Windy} & \textbf{Warm} & \textbf{Cold} \\ \hline
    Tea & \textbf{8.78\%} & \textbf{6.64\%} & \textbf{7.70\%} & \textbf{10.08\%} \\ \hline
    Coffee & \textbf{6.59\%} & \textbf{5.94\%} & \textbf{6.77\%} & \textbf{6.73\%} \\ \hline
    Meat & 4.20\% & \textbf{9.44\%} & 4.38\% & 3.95\% \\ \hline
    Chocolate & \textbf{4.83\%} & 3.50\% & \textbf{4.56\%} & \textbf{5.14\%} \\ \hline
    Cake & 2.77\% & 4.20\% & 2.85\% & 2.93\% \\ \hline
    Ice cream & 3.05\% & 1.75\% & 4.04\% & 2.39\% \\ \hline
    Salad & 2.19\% & 3.15\% & 2.14\% & 1.81\% \\ \hline
    Dumplings & 2.25\% & 1.05\% & 2.28\% & 2.12\% \\ \hline
    Pancake & 2.16\% & 0.70\% & 2.07\% & 2.20\% \\ \hline
    Sauce & 2.01\% & 0.70\% & 2.07\% & 1.65\% \\ \hline
    Gingerbread & 1.49\% & 2.10\% & 0.74\% & 2.10\% \\ \hline
    \end{tabular}
    \caption{Comparison of top products during windy (wind speed $\geq$ 20km/h), rainy (precipitation > 0), cold ($\leq$ 0 $^{\circ}$C), and warm weather ($\geq$ 0 $^{\circ}$C).}
    \label{tab:windy-rainy-cold-warm}
\end{table}

\begin{table}[t]
    \small
    \centering
    \begin{tabular}{|l|c|c|c|}
    \hline
     & \textbf{Negative} & \textbf{Neutral} & \textbf{Positive} \\ \hline
    Cold & 12.59\% & 37.25\% & 50.17\% \\ \hline
    Warm & 13.20\% & 38.68\% & 48.12\% \\ \hline
    Windy & \textbf{23.15\%} & \textbf{48.40\%} & 28.45\% \\ \hline
    Snowy & 11.88\% & 36.06\% & 52.06\% \\ \hline
    Rainy & 13.63\% & 38.64\% & 47.73\% \\ \hline
    High Pres & \textbf{23.10\%} & \textbf{48.26\%} & 28.63\% \\ \hline
    Low Pres & 12.63\% & 38.72\% & 48.65\% \\ \hline \hline
    Overall & 13.07\% & 38.38\% & 48.55\% \\ \hline
    \end{tabular}
    \caption{Weather relation to tweet sentiment.}
    \label{tab:emo-weather}
\end{table}

\section{Conclusion}
\label{sec:conclusion}

This paper contributes to understanding how weather impacts the mood of food consumers by examining influence on food tweets. The knowledge can be useful to public health policymakers and applied when nudging consumers to choose more healthy food alternatives in different weather conditions and seasons. 
Obesity, type 2 diabetes and cardiovascular diseases are just a few of the health problems acquired due to nutritional specifics \cite{Min,Mai}. The global spread of obesity has been labelled a pandemic and it is of utmost importance to understand the underlying factors behind food choice. Acknowledging and understanding the impact of weather on food consumers and their affective reactions helps explain the complexities associated - food waste, healthy vs. unhealthy choices and other issues.

We also highlight the lack of weather data to obtain precise results. A more fine-grained and longitudinal weather data set could allow for higher precision for food tweet data correlation. Besides that, there should also be additional studies done with regard to other cyclical events encountered in modern lives - e.g. school schedule and holidays, annual sport events and others - to capture the impact of weather and non-weather related seasonality on food tweet sentiment.

We aim to contextualise the behaviour of tweeting about food in a given geographical area and build a framework for more nuanced understanding of food-related discourse in Latvian language Twitter \cite{velasco2021gastrophysics}. The contextual knowledge created can be helpful to researchers working with personalised food and health application model development, since humans are social beings, and peer behaviour impacts their choice. Furthermore, we wish to highlight how interconnected our digital and analogue lives are - following up the tweet sentiment and frequency indicators with actual purchasing behaviour and food sales data. We plan to release the tweet-weather dataset as an addition to the existing LTEC and make it public on GitHub.

\bibliography{j_yourrefs,anthology}
\bibliographystyle{acl_natbib} 


\clearpage
\appendix
\onecolumn 
\section{Weather Data Availability}
\begin{figure*}[hb]
  \includegraphics[width=\linewidth]{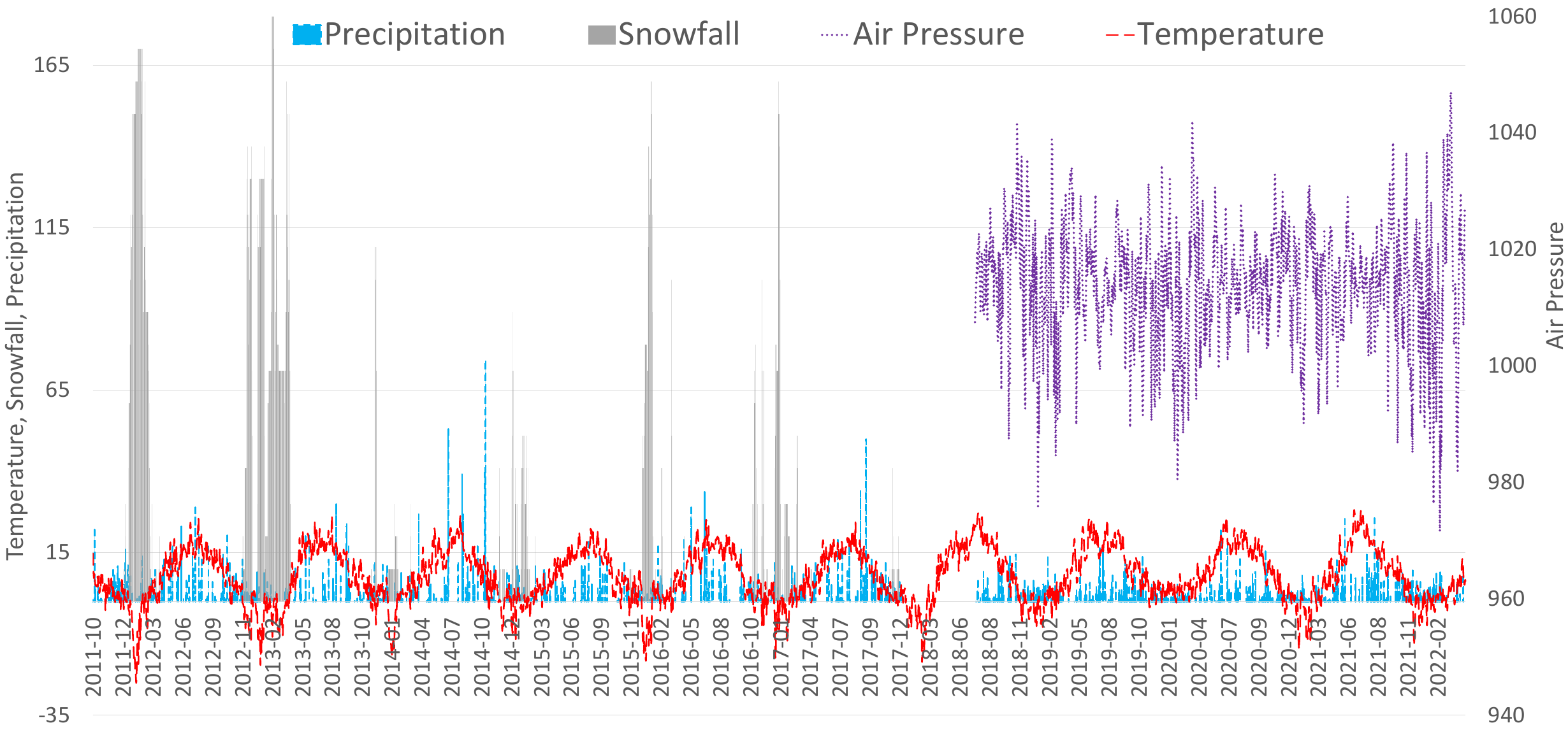}
  \caption{Available weather data from Meteostat.}
  \label{fig:weather-data}
\end{figure*}

Figure \ref{fig:weather-data} shows a visualisation of the data. We could reliably obtain only data for temperature and precipitation for the entire period. There is only a slight gap in precipitation data for the first half of 2018. However, data for snowfall was only available up to February of 2017, and data for wind speed and air pressure was only available from August of 2018 onward. 

\end{document}